\documentclass[11pt]{article}
\usepackage{amssymb,amsmath,amsfonts}
\usepackage{graphicx}
\usepackage{graphics}
\usepackage{eepic,epsfig}
\usepackage{verbatim}
\usepackage{color}
\usepackage{hyperref}
1.60cm \makeatletter \@addtoreset{equation}{section}

\makeatother

\begin{document}
\title{Induced vacuum bosonic current in a compactified cosmic string spacetime}
\author{E. A. F. Bragan\c{c}a$^1$\thanks{E-mail:deduardo@lnf.infn.it} ,
H. F. Santana Mota$^2$\thanks{E-mail: hm288@sussex.ac.uk} and E. R. Bezerra de Mello$^3$\thanks
{E-mail: emello@fisica.ufpb.br}\\
\\
\textit{$^{1,3}$Departamento de F\'{\i}sica, Universidade Federal da Para\'{\i}ba}\\
\textit{58.059-970, Caixa Postal 5.008, Jo\~{a}o Pessoa, PB, Brazil}\\
\vspace{0.1cm}\\
\textit{$^{1}$INFN, Laboratori Nazionali di Frascati,}\\
\textit{Via Enrico Fermi 40, 00044 Frascati, Italy}\\
\vspace{0.1cm}\\
\textit{$^{2}$Department of Physics and Astronomy, University of Sussex}\\
\textit{Falmer, Brighton BN1 9QH, U.K.}}
\maketitle
%
\begin{abstract}
%
We analyze the bosonic current densities induced by a magnetic flux
running along an idealized cosmic string considering that
the coordinate along its axis is compactified. We also consider
the presence of a magnetic flux enclosed by the compactificatified axis.
To develop this analysis, we calculate the complete set of normalized bosonic wave
functions obeying a quasiperiodicity condition along the compactified dimension.
We show that in this context only the azimuthal and axial currents
take place. 
\end{abstract}
\bigskip


\bigskip
%
\section{Introduction}
\label{Int}
%
Cosmic strings are linear topological defects which may have been created
in the early universe as a consequence of phase
transitions and are predicted in the context of the standard gauge field theory of elementary
particle physics \cite{VS,hindmarsh,Hyde:2013fia}. The formation of cosmic string can have
astrophysical and cosmological consequences. For instance, emission of gravitational waves and high energy cosmic
rays by strings such as neutrinos and gamma-rays, along observational data, can help to constraint the product
of the Newton's constant, $G$, and the linear mass density of the string, $\mu_{0}$ \cite{hindmarsh}.

The geometry associated with an infinity and straight cosmic string is locally flat but topologically conical,
having a planar angle deficit given by $\Delta\phi=8\pi G\mu_{0}$ on the two-surface orthogonal to the string.
This conical structure alters the vacuum fluctuations associated with quantum fields and changes the vacuum
expectation values (VEVs) of physical observables like the energy-momentum tensor, $\langle T_{\mu \nu}\rangle$
\footnote{The calculation of the VEV of physical observables associated with scalar and
fermionic fields in the cosmic string spacetime has been developed in  Refs.
 \cite{PhysRevD.35.536,escidoc:153364,GL,DS,PhysRevD.46.1616,
 PhysRevD.35.3779,LB,Moreira1995365,BK}.}.  Moreover, the presence of a magnetic flux running
 along the string's axis provides additional contributions to these VEVs  associated with
 charged fields \cite{PhysRevD.36.3742,guim1994,SBM,SBM2,SBM3,Spinelly200477,
 SBM4,Braganca:2014qma}. Specifically an important quantity induced in this
 context is the current density, $\langle j^\mu\rangle$. 
 
 The presence of compact dimensions also induces topological quantum effects on matter field.
An interesting application of field theoretical models that present compact dimensions can be found
in nanophysics. The long-wavelength description of the electronic states in graphene can be formulated
in terms of the Dirac-like theory in three-dimensional spacetime, with the Fermi velocity playing the
role of the speed of light \cite{RevModPhys.81.109}.
%
\section{Wightman function}
\label{Wightman}
%

Here we consider a $(3+1)-$dimensional cosmic string spacetime. By using  cylindrical
coordinates $(x^{1},x^{2},x^{3})=(r,\phi,z)$ with the string on the $1-$dimensional
hypersurface $r=0$, the corresponding geometry is described by the line element
\begin{equation}
ds^{2}=g_{\mu\nu}dx^{\mu}dx^{\nu}=dt^{2}-dr^2-r^2d\phi^2-dz^2\ .
\label{eq01}
\end{equation}
The coordinates take values in the following intervals: $r\geq 0$, $0\leq\phi\leq 2\pi/q$ and 
$-\infty< t < +\infty$. The parameter $q\geq 1$ codifies the presence of the
cosmic string. Moreover, we assume that the direction
along the $z$-axis is compactified to a circle with length $L$, so $0\leqslant z\leqslant L$.
\footnote{In the standard cosmic string spacetime we have $q^{-1}=1-4\mu_0$,
being $\mu_0$ the linear mass density of the string.}

We are interested in determining the induced vacuum current density, $\langle j_{\mu}\rangle$,
associated with a charged scalar quantum field, in the presence of a magnetic fluxes running along the core of the
string and enclosed by it.  To develop this analyze, we shall calculate the corresponding complete set of normalized
bosonic wave function.

The equation that governs the quantum dynamics of this system is
\begin{equation}
\left[\frac{1}{\sqrt{|g|}}D_\mu\left(\sqrt{|g|}\,g^{\mu\nu}D_\nu\right)+m^{2}+\xi R\right] \varphi (x)=0 \ ,
\label{eq02}
\end{equation}
where $D_{\mu}=\partial_{\mu}+ieA_{\mu}$ and $g=det(g_{\mu \nu})$. We considered the presence
of an non-minimal coupling, $\xi$, between the field and the geometry represented by the Ricci scarlar,
$R$. However, for a thin and infinitely straight cosmic string, $R=0$ for $r\neq0$. To develop
our analysis, will be assumed the $z-$axis is compactified to a circle with lenght $L$: $
0\leqslant z\leqslant L$. The compactification is achieved by imposing the quasiperidicity condition
on the matter field
\begin{equation}
\varphi (t,r,\phi,z+L,x^4)=e^{2\pi i\beta}\varphi(t,r,\phi,z,x^4) \ ,  
\label{eq03}
\end{equation}
with a constant phase $\beta $, $0\leqslant \beta \leqslant 1$. The special
cases $\beta =0$ and $\beta =1/2$ correspond to the untwisted and twisted
fields, respectively, along the $z$-direction. In addition, we shall consider the
presence of the following constant vector potential
\begin{equation}
A_{\mu}=(0,0,A_{\phi},A_{z})\ , 
\label{eq04}
\end{equation}
with $A_{\phi}=-q\Phi_\phi/(2\pi)$ and $A_{z}=-\Phi_z/L$, being $\Phi_\phi$ and $\Phi_z$
the corresponding magnetic fluxes. In quantum field  theory, the  condition
\eqref{eq03} alters the spectrum of the
vacuum fluctuations compared with the case of uncompactified dimension and,
as a consequence, the induced vacuum current density changes.

Using Eqs. \eqref{eq01} and  \eqref{eq02},  and considering the general expression
$\varphi(x)=CR(r)e^{-i\omega t+iqn\phi+ik_z z}$, the normalized expression
takes the form
\begin{equation}
\varphi_{\sigma}(x)=\left[\frac{q\lambda}{4\pi \omega_l L}\right]^{\frac{1}{2}} 
J_{q|n+\alpha|}(\lambda r)e^{-i\omega t+iqn\phi+ik_l z} \   ,
\label{eq05}
\end{equation}
being specified by the set  quantum numbers, $\sigma=(\lambda, k_{z},n)$.
Also, we have
\begin{eqnarray}
\lambda =\sqrt{\omega^2-{\tilde{k}}_z^2-m^2} \ ,  \quad
\alpha =\frac{eA_{\phi}}{q}=-\frac{\Phi_{\phi}}{\Phi_{0}}\ , \quad
\tilde{k_{z}}=k_{z}+eA_{z} \     ,
\label{eq06}
\end{eqnarray}
with $\Phi_{0}=2\pi/e$ being the quantum flux.
As a consequence of the condition \eqref{eq03}, the quantum number $k_z$ is discretized as follows:
\begin{equation}
k_z=k_l=\frac{2\pi}{L}(l+\beta) \ \ {\rm with} \  l=0\ ,\pm 1,\pm 2,... \ .
\label{eq07}
\end{equation}
And under this circumstancy, the energy takes the form
\begin{equation}
\omega=\omega_l=\sqrt{m^2+\lambda^2+{\tilde{k}}^2_l} \ , 
\label{eq08}
\end{equation}
where
\begin{eqnarray}
{\tilde{k}}_z={\tilde{k}}_l=\frac{2\pi}{L}(l+\tilde{\beta}) \ , \quad
\tilde{\beta}=\beta+\frac{eA_zL}{2\pi}=\beta-\frac{\Phi_z}{\Phi_0} \ .
\label{eq09}
\end{eqnarray}

The properties of the vacuum state are described by the corresponding positive 
frequency Wightman function, $W(x,x')=\left\langle 0|\varphi(x)\varphi^{*}(x')|0 \right\rangle$, 
where $|0 \rangle$ stands for the vacuum state.  Having the complete set of normalized mode functions, 
$\{\varphi_{\sigma}(x), \ \varphi_{\sigma}^{*}(x')\}$, satisfying the periodicity condition \eqref{eq03}, 
we can evaluate the corresponding  Wightman function as:
\begin{equation}
W(x,x')=\sum_{\sigma}\varphi_{\sigma}(x)\varphi_{\sigma}^{*}(x')\quad \rm{with} \quad
\sum_{\sigma }=\sum_{n=-\infty}^{+\infty} \ \int_0^\infty
\ d\lambda  \sum_{l=-\infty }^{+\infty} \ 
\label{eq10}
\end{equation}
The  mode functions in Eq. \eqref{eq05} are specified by the set of quantum numbers 
$\sigma=(n,\lambda,k_l)$, with the values in the ranges $n=0,\pm1,\pm2, \ \cdots$,  $0<\lambda<\infty$ and $k_l=2\pi(l+\beta)/L$
with $l=0,\pm1,\pm2, \ \cdots$. 

Substituting \eqref{eq05} into the sum \eqref{eq10} we obtain
\begin{eqnarray}
W(x,x')&=&\frac{q}{4\pi L}\sum_\sigma e^{iqn\Delta\phi} 
\lambda J_{q|n+\alpha|}(\lambda r) J_{q|n+\alpha|}(\lambda r')
\frac{e^{-i\omega_l\Delta t+ik_l\Delta z}}{\omega_l} \ ,
\label{eq11}
\end{eqnarray}
where $\Delta\phi=\phi-\phi'$, $\Delta t=t-t'$ and $\Delta z=z-z'$. 

Now, we are in position
to calculate the induced vacuum  bosonic current density, $\langle j_\mu\rangle$.
This calculation will be developed in the next section.
%
\section{Bosonic current}
\label{current}
%
The bosonic current density operator is given by,
\begin{eqnarray}
j_{\mu }(x)&=&ie\left[\varphi ^{*}(x)D_{\mu }\varphi (x)-
(D_{\mu }\varphi)^{*}\varphi(x)\right] \nonumber\\
&=&ie\left[\varphi^{*}(x)\partial_{\mu }\varphi (x)-\varphi(x)
(\partial_{\mu }\varphi(x))^{*}\right]-2e^2A_\mu(x)|\varphi(x)|^2 \   .
\label{eq12}
\end{eqnarray}
Its vacuum expectation value (VEV) can be evaluated in terms of the positive frequency Wightman 
function as shown below:
\begin{equation}
\left\langle j_{\mu}(x) \right\rangle=ie\lim_{x'\rightarrow x}
\left\{(\partial_{\mu}-\partial_{\mu}')W(x,x')+2ieA_\mu W(x,x')\right\} \ .
\label{eq13}
\end{equation}
Writing  the  parameter $\alpha$ in Eq. \eqref{eq06} in the following form
\begin{equation}
\alpha = n_{0}+\alpha_{0}  \ {\rm with} \ |\alpha_{0}|<\frac{1}{2} \ ,
\label{alphazero}
\end{equation}
where $n_{0}$ is an integer number, we  can see that the above  VEV is a periodic 
function of the magnetic fluxes $\Phi_{\phi}$ and $\Phi_{z}$
with period equal to the quantum flux.  In fact, as we will see, the VEV of the current density
depends on $\alpha_{0}$ only.

In the next section we will calculate the current densities. It has been shown 
in Ref. \cite{Braganca:2014qma},  that the charge density and the radial current density 
vanish for the system in consideration. So,  here we will focus only
in the calculations of the azimuthal and axial current densities.

%
\subsection{Azimuthal current}
%
The VEV of the azimuthal current density  is given by:
\begin{equation}
\left\langle j_{\phi}(x) \right\rangle = ie \lim_{\phi '\rightarrow \phi}
\left\{(\partial_{\phi}-\partial_{\phi '})W(x,x')+2ieA_{\phi}W(x,x')\right\} \ .
\label{jphi}
\end{equation}
Substituting \eqref{eq11} into the above equation, 
we get the formal expression for the azimuthal bosonic current density below,
\begin{eqnarray}
\left\langle j_{\phi}(x) \right\rangle&=&-\frac{qe}{4\pi L}\sum_{n=-\infty}^\infty
q(n+\alpha)\int_0^\infty \ d \lambda \ \lambda \ J^2_{q|n+\alpha|}(\lambda r)
\sum_{l=-\infty}^\infty\frac{1}{\sqrt{m^2+\lambda^2+{\tilde{k}}^2}}. \nonumber\\
\label{jphi1}
\end{eqnarray}

To develop the summation over the quantum number $l$ we shall apply
the Abel-Plana summation formula in the form used in  Ref. \cite{PhysRevD.82.065011},
which is given by
\begin{eqnarray}
&&\sum_{l=-\infty }^{\infty }g(l+\tilde{\beta} )f(|l+\tilde{\beta} |)=\int_{0}^{\infty }du\,
\left[ g(u)+g(-u)\right] f(u)  \notag \\
&&\qquad +i\int_{0}^{\infty }du\left[ f(iu)-f(-iu)\right] \sum_{\lambda =\pm
1}\frac{g(i\lambda u)}{e^{2\pi (u+i\lambda \tilde{\beta} )}-1} \ .
\label{sumform}
\end{eqnarray}
Taking $g(u)=1$ and 
\begin{equation}
f(u)=\frac{1}{\sqrt{(2\pi u/L)^{2}+\lambda^2+m^{2}}} \   , 
\label{fg}
\end{equation}
it is possible to decompose the expression
to $\langle j_{\phi}\rangle$ as the sum of the
two contributions as shown below:
\begin{equation}
\langle j_{\phi}\rangle =\langle j_{\phi }\rangle _{{\rm cs}}+\langle j_{\phi
}\rangle _{{\rm c}} \   .  
\label{total}
\end{equation}
The first contribution, $\langle j_{\phi }\rangle _{{\rm cs}}$ corresponds
to the azimuthal current density in the geometry of the 
3-dimensional cosmic string spacetime without  
compactification. It is given by the first integral of (\ref{sumform}). 
The second term, $\langle j_{\phi }\rangle _{{\rm c}}$, is induced by the
compactification of the string along its axis and is provided by the second integral.

For the frist contribution we have:
\begin{equation}
\left\langle j_{\phi}(x) \right\rangle _{{\rm cs}}= -\frac{eq}{2\pi^{2}} 
\sum_{n=-\infty}^{\infty}q(n+\alpha)\int_{0}^{\infty}d\lambda \ 
\lambda J_{q|n+\alpha|}^{2}(\lambda r)\int_{0}^{\infty}\frac{dy}{\sqrt{y^{2}+
\lambda^{2}+m^{2}}} \ ,
\label{jphi_cs02}
\end{equation}
where we have introduced a new variable $y=2\pi u/L$. Using the identity
\begin{equation}
\frac{1}{\sqrt{m^2+\lambda^2+{\tilde{k}}^2}}=\frac{2}
{\sqrt{\pi}}\int_{0}^{\infty}ds \ e^{-(m^2+\lambda^2+{\tilde{k}}^2)s^{2}} \ ,
\label{ident}
\end{equation}
the integral over $\lambda$ can be evaluated by using Ref. \cite{gradshteyn2000table}.
Finally, writing $\alpha$ in the form \eqref{alphazero} we obtain
\begin{equation}
\left\langle j_{\phi}(x) \right\rangle _{{\rm cs}}=-\frac{eq^2}{(2\pi)^{2}r^{2}}\int_{0}^{\infty}
dw \ e^{-w-\frac{m^{2}r^{2}}{2w}}\sum_{n=-\infty}^{\infty}(n+\alpha_{0})
I_{q|n+\alpha_{0}|}(w) \ ,
\label{jphi_cs03}
\end{equation}
where we have defined $w=r^2/2s^2$. Using the result for the summation over $n$ found
in Ref. \cite{Braganca:2014qma}, we obtain
\begin{eqnarray}
\left\langle j_{\phi}(x) \right\rangle_{{\rm cs}}&=&-\frac{em^{4}r^2}{\pi^{2}}
\left[\sideset{}{'}\sum_{k=1}^{[q/2]}\sin(2k\pi/q)\sin(2k\pi\alpha_{0})
 f_{2}\left[2mr\sin(k\pi/q)\right]\right.\nonumber\\
&+&\left.\frac{q}{\pi}
\int_{0}^{\infty}dy \ \frac{g(q,\alpha_{0},2y)\sinh (2y)}{\cosh(2qy)-\cos(q\pi)}
f_{2}\left[2mr\cosh(y)\right]
\right],
\label{jphi_cs04}
\end{eqnarray}
where we use the notation
\begin{equation}
f_{\nu}(x)=\frac{K_{\nu}(x)}{x^{\nu}} \ .
\end{equation}
In the above equation
\begin{eqnarray}
g(q,\alpha_0,2y)&=&\sin(q\pi\alpha_0)\sinh[(1-|\alpha_0|)2qy]
-\sinh(2qy\alpha_0)\sin[(1-|\alpha_0|)\pi q].
\label{g_function}
\end{eqnarray}

\begin{figure}[htb]
\centerline
{\includegraphics[width=0.4\textwidth]{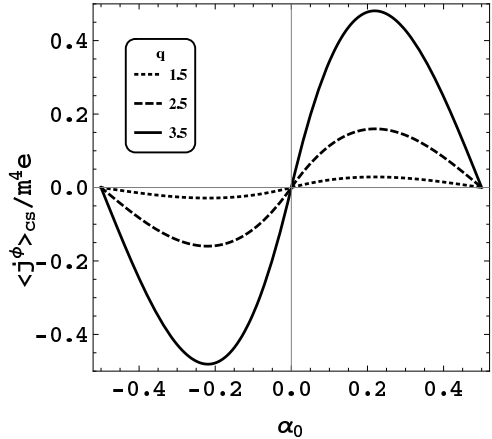}}
\caption{The azimuthal current density without compactification
is plotted, in units of ``$m^4e$'', in terms of $\alpha_0$ for $mr=0.5$ and $q=1.5, 2.5$ and $3.5$.}
\label{fig1}
\end{figure}

We can see that where $\alpha_{0}=0$, $\left\langle j_{\phi}(x) \right\rangle_{{\rm cs}}$
vanishes. From Eq. \eqref{jphi_cs04}, we can see that $\left\langle j_{\phi}(x) \right\rangle_{{\rm cs}}$
is an odd function of $\alpha_{0}$, with period equal to the quantum flux $\Phi_0$.
We plot in Fig.\ref{fig1} the behavior of the azimuthal current density as function of $\alpha_{0}$
for specific values of the parameter $q$, considering $mr=0.5$. 
From the figure is possible to conclude
that the main influence of this parameter is amplify the oscilatory nature of the azimuthal current with
respect to $\alpha_{0}$.

At large distances from the string, $mr\gg1$, and considering $q>2$, the
azimuthal current due the cosmic string is dominated by the first term of 
\eqref{jphi_cs04}, with $k=1$ and is given by
\begin{eqnarray}
\left\langle j_{\phi}(x) \right\rangle_{{\rm cs}}&\approx& -\frac{em}{(2\pi)^2}
\left(\frac m{\pi r \sin(\pi/q)}\right)^{\frac{1}{2}}
\sin(2\pi\alpha_{0})
 \cot(\pi/q)e^{-2mr\sin(\pi/q)} \ ,
\label{jphi_csmrlarged3}
\end{eqnarray}
where we see an exponential decay.

To obtain the contribution for the azimuthal current induced by the compactification
we substitute the second term of \eqref{sumform} into Eq. \eqref{jphi1}.
Also using the series expansion $(e^{u}-1)^{-1}=\sum_{l=1}^{\infty}e^{-lu}$
and the following representation for the Macdonald function
 \cite{gradshteyn2000table}
\begin{equation}
K_{\nu}(x)=\frac{1}{2}\left(\frac{x}{2}\right)^{\nu}\int_{0}^{\infty}
dt\frac{e^{-t-\frac{x^{2}}{4t}}}{t^{\nu +1}}\    ,
\label{Macdonald_repres}
\end{equation}
we obtain
\begin{eqnarray}
\left\langle j_{\phi}(x) \right\rangle_{{\rm c}}&=&-\frac{eq^2}{\pi r^2}
\sum_{l=1}^{\infty}\cos(2\pi l\tilde{\beta})
\int_{0}^{\infty}dw \
e^{-w\left[1+\frac{l^{2}L^{2}}{2r^{2}}\right]-\frac{r^{2}m^{2}}{2w}}
\sum_{n=-\infty}^{\infty}(n+\alpha_{0})
I_{q|n+\alpha_{0}|}(w), \nonumber\\
\label{jphi_comp03}
\end{eqnarray}
where we have used $\alpha$ in the form \eqref{alphazero} and defined $w=\frac{2r^{2}t}{l^{2}L^{2}}$.
Using the result  for the summation on $n$ found in Ref. \cite{Braganca:2014qma} in
the above equation the final expression for the azimuthal current induced by the compactification is
\begin{eqnarray}
\left\langle j_{\phi}(x) \right\rangle_{{\rm c}}&=&-\frac{2em^{4}r^2}{\pi^{2}}
\sum_{l=1}^{\infty}
\cos(2\pi l\tilde{\beta})\left\{\sideset{}{'}\sum_{k=1}^{[q/2]}\sin(2k\pi/q)
\sin(2k\pi\alpha_{0})f_{2}\left[mL\sqrt{l^{2}+\rho_{k}^{2}}\right]\right.\nonumber\\
&&\left.+\frac{q}{\pi}\int_{0}^{\infty}dy\frac{g(q,\alpha_{0},2y) \sinh{(2y)}
 }{\cosh(2qy)-\cos(q\pi)}f_{2}\left[mL\sqrt{l^{2}+\eta^{2}(y)}\right] \right\} \ ,
\label{jphi_comp04}
\end{eqnarray}
where we have defined
\begin{eqnarray}
\rho_{k}=\frac{2r\sin(k\pi/q)}{L} \ , \ \eta(y)=\frac{2r\cosh (y)}{L} \ .
\label{rhoandeta}
\end{eqnarray}

\begin{figure}[htb]
\centerline{
\includegraphics[width=0.4\textwidth]{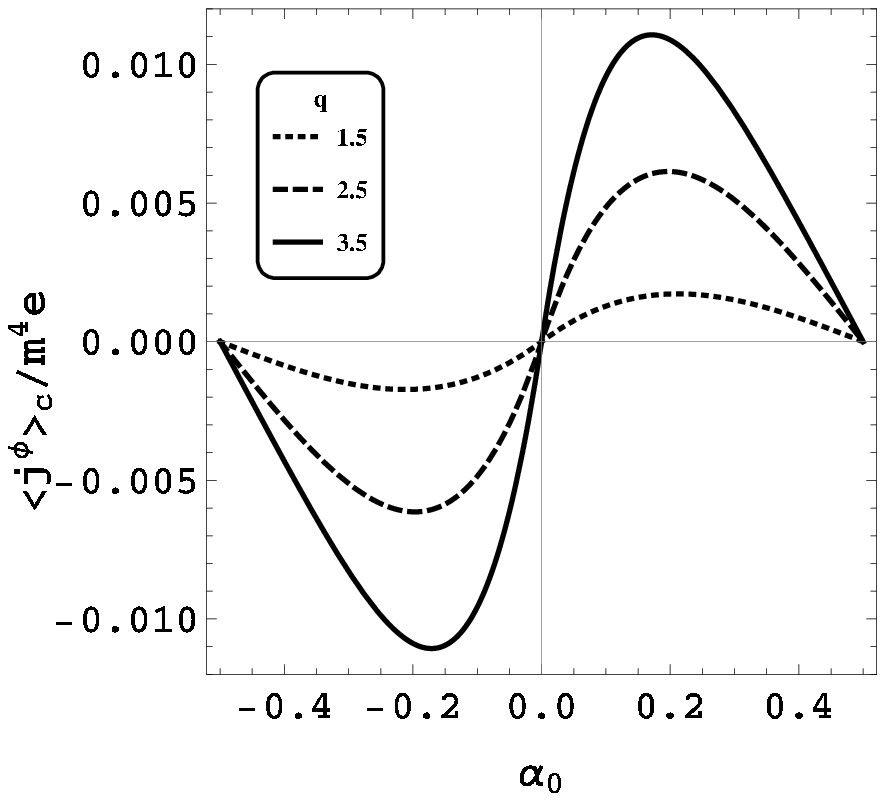}
\includegraphics[width=0.4\textwidth]{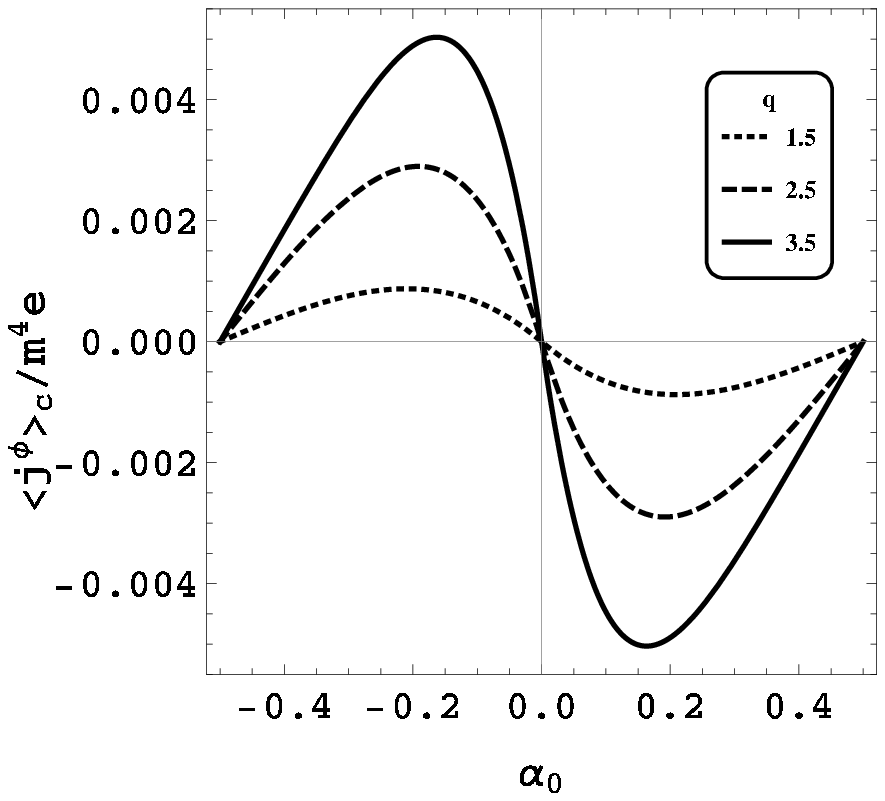}}
\caption{$\left\langle j_{\phi}(x) \right\rangle_{{\rm c}}$
is plotted in units of ``$e/L^2$'',
in terms of $\alpha_0$ for $mr=0.5$, $mL=1$ and $q=1.5, 2.5$ and 3.5.
The plot on the left is for $\tilde{\beta}=0.1$ while the plot on the
right is for $\tilde{\beta}=0.7$.}
\label{fig2}
\end{figure}

From  Eq. \eqref{jphi_comp04} is possible to see that the contribution induced by
the compactification for the azimuthal current is an even function of the parameter $\tilde{\beta}$ and
is an odd function of the magnetic flux along the core of the string, with period equal to the quantum flux.
This contribution vanishes for the case where $\alpha_{0}=0$. In Fig. \ref{fig2} we plot
the behavior of the compactified azimuthal current density as a function of $\alpha_{0}$
for different values of the parameter $q$ and considering $mr=0.5$ and $mL$=1. Is possible to see that
effect of the parameter $q$ is increase the oscillatory nature of the azimuthal current while the parameter
$\tilde{\beta}$ changes the direction of oscillation.

For large values of the length of the compact dimension $mL\gg1$, assuming 
that $mr$ is fixed, the main contribution comes from the $l=1$ term . It
is given by:
\begin{eqnarray}
\left\langle j_{\phi}(x) \right\rangle_{{\rm c}}&\approx &-\frac{\sqrt{2} e r^{2}m^{\frac{3}{2}}
\cos(2\pi\tilde{\beta})e^{-mL}}{\pi^{\frac{9}{2}}L^{\frac{5}{2}}}\left[
\sideset{}{'}\sum_{k=1}^{[q/2]}\sin(2k\pi/q)\sin(2k\pi\alpha_{0})\right.\nonumber \\
&+&\left. \frac{q}{\pi}\int_{0}^{\infty}dy\frac{g(q,\alpha_{0},2y) \sinh (2y)}
{\cosh(2qy)-\cos(q\pi)}\right] \ ,
\end{eqnarray}
where we can see an exponential decay. This means that  in this limit,
the contribution for the total  azimuthal current density  is dominated by $\langle j_\phi(x)\rangle_{{\rm cs}}$.
%
\subsection{Axial current}
\label{axial}
%
The VEV of the axial current is given by
\begin{equation}
\left\langle j_{z}(x) \right\rangle = ie\lim_{z'\rightarrow z}
\left\{(\partial_{z}-\partial_{z'})W(x,x')+2ieA_{z}W(x,x')\right\} \ .
\label{jaxial}
\end{equation}
Using the fact that $A_z=-\Phi_z/L$, a formal expression for this current can be provided.
It reads,
\begin{equation}
\left\langle j_{z}(x) \right\rangle = -\frac{eq}{2\pi L}\sum_{n=-\infty}^\infty
\int_0^\infty \lambda \ J^2_{q|n+\alpha|}(\lambda r) 
\sum_{l=-\infty}^\infty\frac {{\tilde{k}}_l}{\sqrt{m^2+\lambda^2+{\tilde{k}}_l}} \ ,
\label{jaxial1}
\end{equation}
where ${\tilde{k}}_l$ is given by \eqref{eq06}. To evaluate the summation over the quantum number $l$
we use again  Eq. \eqref{sumform}. For this case we have $g(u)=2\pi u/L $ and $f(u)$
is given by Eq. \eqref{fg}. Due the fact the $g(u)$ is an odd function, the first term on the
right-hand side of \eqref{sumform} vanishes. Thus, the only contribution for the axial
current is due the second term of \eqref{sumform}, that means that the axial current
is due only to  the compactification. Adopting similar steps that we made to derivate
 $\left\langle j_{z}(x) \right\rangle_{cs}$,
for this case we obtain:
\begin{eqnarray}
\left\langle j_{z}(x) \right\rangle_{{\rm c}}=\frac{qeL}{2\pi^{2}r^{4}}
\sum_{l=1}^{\infty}l\sin(2\pi l\tilde{\beta})\int_{0}^{\infty}dw \ w
e^{-w\left[1+\frac{l^2L^2}{2r^2}\right]-\frac{m^2r^2}{2w}}
\sum_{n=-\infty}^{\infty}I_{q|n+\alpha_{0}|}(w), \nonumber\\
\label{jaxialc1}
\end{eqnarray}
where we have defined the variable $w$ as the same as in \eqref{jphi_comp03}. The summation
on $n$ in the above equation also was developed in  Ref. \cite{Braganca:2014qma}.
Using that result, is possible to write the axial current as follows
\begin{eqnarray}
\left\langle j_{z}(x) \right\rangle_{{\rm c}}=\frac{em^2}{\pi^{2}L}
\sum_{l=1}^{\infty}\frac{\sin(2\pi l\tilde{\beta})}{l}K_{2}(lmL)\
+\ \left\langle j_{z}(x) \right\rangle_{{\rm c}}^{(q,\alpha_{0})}.
\end{eqnarray}
The first contribution in the right-hand side of the above equation
is independent of the $\alpha_{0}$ and the parameter $q$. It is a pure
topological term, induced by the compactification only. The second
contribution comes from the magnetic flux and
planar angle deficit. It is:
\begin{eqnarray}
\left\langle j_{z}(x) \right\rangle_{{\rm c}}^{(q,\alpha_{0})}&=&\frac{2em^{4}L}
{\pi^{2}}\sum_{l=1}^{\infty}l\sin(2\pi l\tilde{\beta})
\left\{\sideset{}{'}\sum_{k=1}^{[q/2]}\cos(2k\pi\alpha_{0})
f_{2}\left[mL\sqrt{l^{2}+\rho_{k}^{2}}\right]\right.\nonumber\\
&-&\left.\frac{q}{\pi}\int_{0}^{\infty}dy
\frac{f(q,\alpha_{0},2y)}{\cosh(2qy)-\cos(q\pi)}
f_{2}\left[mL\sqrt{l^{2}+\eta^{2}(y)}\right]\right\}
\label{jaxial2_comp}
\end{eqnarray}
where $\rho_k$ and $\eta(y)$ are given by \eqref{rhoandeta} and
\begin{eqnarray}
f(q,\alpha_0,2y)=\sin[(1-|\alpha_0|)\pi q]\cosh(|\alpha_0| 2qy)
+\sin(|\alpha_0|\pi q)\cosh[(1-|\alpha_0|)2qy].
\end{eqnarray}

\begin{figure}[htb]
\centerline{\includegraphics[width=0.4\textwidth]{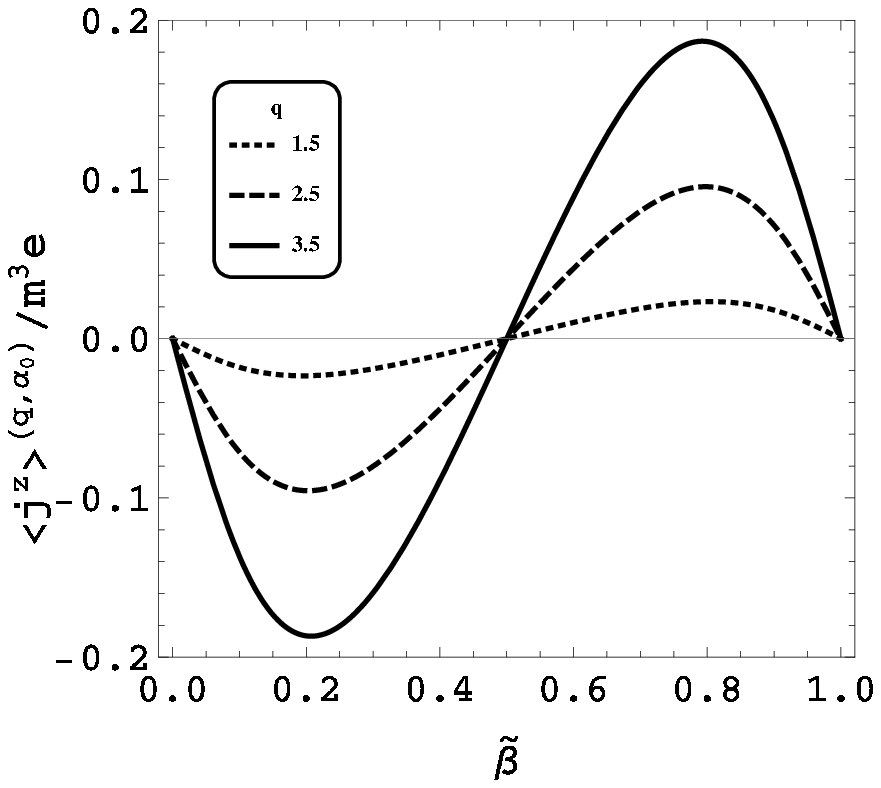}
\includegraphics[width=0.4\textwidth]{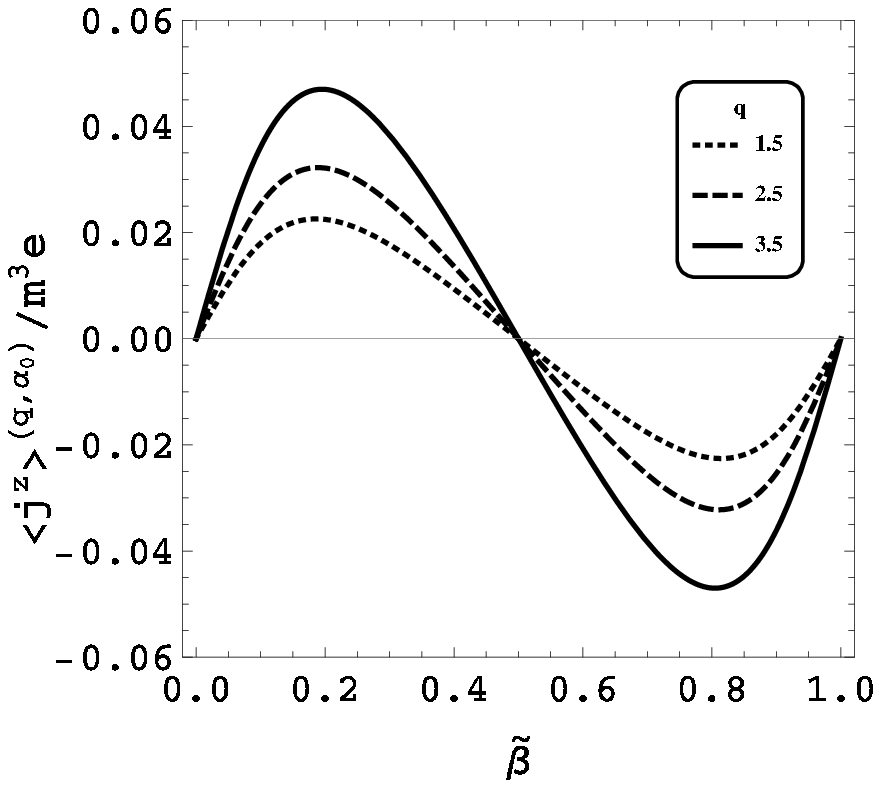}}
\caption{$\left\langle j_{z}(x) \right\rangle_{{\rm c}}^{(q,\alpha_{0})}$ is plotted,
in units of ``$m^3e$'', in terms of $\tilde{\beta}$ for the values $mr=0.4$,
$mL=1$ and $q=1.5, 2.5, 3.5$. The left plot is for $\alpha_0=0$ while
the right plot is for $\alpha_0=0.25$.}
\label{fig03}
\end{figure}

The Eq. \eqref{jaxial2_comp} is an odd function of the parameter $\tilde{\beta}$ and is
an even function of $\alpha_{0}$, with period equal to the quantum flux. Is possible to see that this
contribution vanishes for the case where $q=1$ and $\alpha_{0}=0$. In Fig. \ref{fig03}, we plot
the behavior of this expression as function of $\tilde{\beta}$ taking $mr=0.4$, $mL=1$ and $q=1.5, \ 2.5, \ 3.5$.
By these plots we can see
that the amplitude of the current increases with $q$ and the effect of $\alpha_0$
is to change the orientation
of the current.

Considering $mL\gg1$ and $mr$ fixed, the main contribution to \eqref{jaxial2_comp}
comes from the $l=1$ term. So, we have
\begin{eqnarray}
\left\langle j_{z}(x) \right\rangle_{{\rm c}}^{(q,\alpha_{0})}& \approx &
\frac{8em^{\frac{3}{2}}\sin(2\pi\tilde{\beta})e^{-mL}}{(2\pi L)^{\frac{3}{2}}}\left\{
\sideset{}{'}\sum_{k=1}^{[q/2]}\cos(2k\pi\alpha_{0})
- \frac q\pi\int_{0}^{\infty}dy\frac{f(q,\alpha_{0},2y)}{\cosh(2qy)-\cos(q\pi)}\right\},\nonumber\\
\end{eqnarray}
%
\section{Conclusion}
\label{conc}
%
Here, we have investigated the induced bosonic current density in a compactified cosmic string spacetime,
considering the presence of the magnetic fluxes on the azimuthal and axial directions. In order to do that, we
imposed a quasiperiodicity condition, with arbitrary phase $\tilde{\beta}$, on the solution of the
Klein-Gordon equation. We constructed the positive frequency Wightman function \eqref{eq11}, that is necessary
to calculate the induced bosonic current.
We have seen that the compactification induces the azimuthal current density to decompose in two parts.
The first one corresponds to the expression in the geometry of a cosmic string without compactification,
while the second one is due the compactification.
The first contribution is an odd function on the parameter $\alpha_{0}$, with period the quantum flux, and the
second contribution in an even function on the parameter $\tilde{\beta}$. Both contributions vanish
for the case where the parameter $\alpha_{0}$ is equal to zero. In the limit of large values of the parameter
of the compactification, the main contribution come from the first contribution.

We have also shown that the VEV of the axial current density has a purely topogical origin.
This VEV can be expressed as the sum of two terms. One of them is given by the Eq. \eqref{jaxial1}
and is independent of the radial distance $r$, the cosmic string parameter $q$ and the $\alpha_{0}.$
The other contribution is given by the Eq. \eqref{jaxial2_comp} and is due to the magnetic fluxes
and the planar angle deficit. This contribution is an odd function of the parameter $\tilde{\beta}$
and is an even function of the parameter $\alpha_{0}$.%
\section*{Acknowledgments}
The author E. A. F. B thanks the Brazilian agency CAPES and the INFN for the financial support. H. F. S. M thanks
the Brazilian agency CAPES for the financial support. E. R. B. M thanks Conselho Nacional de Desenvolvimento
Cientif\'{i}co e Tecnol\'{o}gico (CNPq) for the partial financial support.
%

\providecommand{\href}[2]{#2}\begingroup\raggedright\endgroup

\end{document}